# Phase separation and phase transitions in undoped and $Rh^{3+}$ doped iron pnictide $CaFe_2As_2$: a Raman scattering study

V. P. Gnezdilov[1,2], A. Yu. Glamazda[1], P. Lemmens[2,3], K. Kudo[4], and M. Nohara[5]

[1] B. Verkin Institute for Low Temperature Physics and Engineering of the National Academy of Sciences of Ukraine Kharkiv 61103, Ukraine
[2] Institute for Condensed Matter Physics, TU-Braunschweig, 38106 Braunschweig, Germany
[3] Laboratory for Emerging Nanometrology and International Graduate School of Metrology, TU-Braunschweig, 38106 Braunschweig, Germany
[4] Department of Physics, Graduate School of Science, Osaka University, Toyonaka, Osaka 560-0043, Japan
[5] Department of Quantum Matter, AdSE, Hiroshima University, Higashi-Hiroshima 739-8530, Japan

*Contact e-mail*: gnezdilov@ilt.kharkov.ua

Iron-pnictides $Ca(Fe_{1-x}Rh_x)_2As_2$ ($x$ = 0, 0.035, and 0.19) were studied across the tetragonal-orthorhombic and uncollapsed → collapsed tetragonal phase transitions using Raman spectroscopy. The effect of the phase separation was observed in the high-temperature phase for the first time. Two phases with the low-spin and high-spin states of the $Fe^{2+}$ ions coexist in the undoped ($x$ = 0) and doped ($x$ = 0.035) samples at ambient conditions. These two phases are present on a larger length scale and they are not limited to the local distortions. Both phases in the undoped sample successfully undergo tetra-to-ortho phase transitions approximately at the same temperature $T_O$ = 160 K. In doped samples, a cascade of phase transitions is observed at temperatures $T^*$ = 100 K and $T_c$ = 72 K with cooling for the first time. A complex type of the re-entrant magnetic structure is realized in the $x$ = 0.035 sample at temperatures below $T_c$, which is distinct from the usual orthorhombic twofold one. The overdoped sample ($x$ = 0.19) also shows phase separation at temperatures 7 – 295 K in Raman spectra similar to these of the remnant phase.

## INTRODUCTION

Starting from the discovery of superconductivity in $LaFeAsO_{1-x}F_x$ with critical temperature, $T_c$ ~ 26 K in 2008 [1], several families of Fe-based superconducting materials have been discovered such as 11-$FeX_{1-x}$ ($X$ = Se/Te/S), 1111-$Re$FeAsO ($Re$ = La/Ce/Pr/Nd/Sm/Gd/Tb) [2], 111-$A$FeAs ($A$ = Li/Na/La) [3], ternary 122-$Ae$Fe$_2$As$_2$ ($Ae$ = Ba/Sr/Ca/Eu) and 122-$Ax$Fe$_2$Se$_2$ ($Ax$ = K/Cs/Rb) [4], 1144-$AeA$Fe$_4$As$_4$ [5], and 42622- $Ae_4M_2O_6Fe_2As_2$ (M = V/Cr/Sc) [6]. All iron-based superconductors consist of $Fe_2X_2$ (X = pnictide/chalcogenide) layers that are composed of edge-sharing, distorted $FeX_4$ tetrahedra. In most cases, these tetrahedra alternate with other intermediate layers [7]. In particular, the combination of effective PtAs layers with FeAs layers led to the discovery of 10-3-8-$Ca_{10}(Pt_3As_8)(Fe_2As_2)_5$ and 10-4-8-$Ca_{10}(Pt_3As_8)(Fe_2As_2)_5$ superconductors [8,9,10]. The high interest and intensive studies of iron-based superconductors quickly led to the appearance of a large number of articles and reviews (see for example, Refs. 11,12,13,14) in which structural features related to superconductivity, doping/pressure-induced superconductivity, and the interplay between magnetism and superconductivity are discussed and summarized.

The 122-$Ae$Fe$_2$As$_2$ materials are among the most interesting materials in the Fe-As family of superconductors with very fragile ground states and extreme sensitivity to external pressure and chemical doping. They are considered to be an ideal platform to study magnetic fluctuation driven superconductivity. It was supposed that all the 122-$Ae$Fe$_2$As$_2$ compounds crystallize in the tetragonal ThCr$_2$Si$_2$-type structure at room temperature. They all exhibit a structural transition upon cooling to an orthorhombic lattice ($T_O$ is ~171 K for Ca [15], ~205 K for Sr [16], ~140 K for Ba [17] and ~200 K for Eu [18]). The structural transition is accompanied by an antiferromagnetic (AFM) ordering of the Fe moments with a wave vector $Q = [1, 0, 1]$ for the spin-density-wave (SDW) pattern. Suitable substitution on either the $Ae$ site or the Fe site can suppress magnetic ordering, and then the system becomes superconducting for certain ranges of doping [19,20,21]. Superconductivity can also be induced in "undoped" and "under-doped" compounds by applying pressure [22].

Generally speaking, the main trends in the study of iron-based compounds are quite similar to copper-based compounds. Namely, the improvement of sample quality as well as unprecedented physical properties became highlighted and opened. Among them:

i) Unusual structural instabilities in 122-$Ae$Fe$_2$As$_2$ compounds which are related to the existence of many energy minima in their energy landscape. CaFe$_2$As$_2$ is conceivably one of the most interesting materials in the family of Fe-pnictide-based superconductors with a very fragile ground state. It combines the salient physical features of the 122-$Ae$Fe$_2$As$_2$ compounds and serves as a readily accessible model system. On the other hand, studies with partly contrasting results lead to several warnings about the complexities inherent in these amazing compounds. In particular, it is well-known now that bulk properties of CaFe$_2$As$_2$ are significantly dependent on the sample's preparation techniques (growth from Sn flux or FeAs self-flux), quenching-temperature, or post-growth treatment [23].

ii) Pressure- or doping-induced isostructural (uncollapsed-tetragonal (ucT) to collapsed-tetragonal (cT)) phase transition in the 122-$Ae$Fe$_2$As$_2$ family lead to an approximately 10% shrinkage of the unit cell's $c$-axis [24,25]. But what is more important, magnetic order and superconductivity are both suppressed. Meanwhile, a Fermi-liquid electronic transport is restored. Interestingly, in $Rh^{3+}$-doped $CaFe_2As_2$ superconductivity occurs in a very narrow range of the Rh-content ($0.020 < x < 0.024$) [26]. It occurs in between the orthorhombic AFM phase and the tetragonal nonmagnetic phase. One should note, that the absence of magnetic order means the absence of a magnetic moment on the iron site. This means that iron ions can be described by a $S = 0$ low spin state [27]. Thus the question arises again about the role of the magnetic fluctuations in the emergence of superconductivity and the orbital or magnetic origin of the ortho-to-tetra structural phase transitions in the $Ca(Fe_{1-x}Rh_x)_2As_2$.

iii) Re-entrance and competition of magnetic phases with orthorhombic $C_2$ and tetragonal $C_4$ rotational symmetry and superconductivity in the region of the superconductivity dome in the phase diagram [28,29,30].

Motivated by the above reasons, we apply Raman spectroscopy to study the parent and $Rh^{3+}$-doped $Ca(Fe_{1-x}Rh_x)_2As_2$ ($x = 0$, 0.035, and 0.19) single crystals. As these stoichiometries represent specific regimes of the phase diagram [26] a focus on the detailed analysis of the temperature-dependent polarized Raman spectra should lead to conclusive results. In spite of the intriguing properties of undoped and doped $CaFe_2As_2$, the number of spectroscopic studies has been limited. Raman spectroscopy is a highly sensitive and informative method that allows to reveal the local crystal distortions, defects, impurities and to probe the phonon, electronic and magnetic excitations simultaneously. As an example, we can cite a series of very successful fundamental Raman studies on the spin and charge degrees of freedom in iron-containing pnictides and chalcogenides [31].

The performed data analysis indicates that in the undoped $CaFe_2As_2$ sample four modes instead of the two symmetry-allowed ones are observed. They are pairs of $A_{1g}$ and $B_{1g}$ modes that give evidence for the presence of at least two tetragonal phases in the sample. This suggests a phase separation into tetragonal uncollapsed and collapsed phases in undoped $CaFe_2As_2$ at ambient conditions. Within the spin state scenario of $Fe^{2+}$ ions, this corresponds to high- and low-spin states, respectively. Both these phases undergo tetra-to-ortho structural phase transitions under cooling. Within the accuracy of our technique, we can suggest a coincidence of temperature scales of these transitions for both phases.

For the $x = 0.035$ sample, we also observed spectra corresponding to two phases at room temperature. However, up to $T = 100$ K we did not observe the appearance of all four lines in the $YX$-scattering geometry. Below some critical $T^* \sim 100$ K we observe the emergence of new phonon lines in coexistence with the previous ones. The only explanation for this feature is the emergence of a new order parameter with different translation symmetry. Upon further cooling to the temperature of the uncollapsed → collapsed tetragonal phase transition $T_{cT}$, cardinal changes occur in the Raman spectra. One can suppose that the last effect is related to the remnant magnetic ordering which should be realized in the uncollapsed phase of the phase-separated $x = 0.035$ sample with iron atoms in a nonzero spin state.

The overdoped sample with $x = 0.19$ located on the right side of the phase diagram [26] of Ca(Fe$_{1-x}$Rh$_x$)$_2$As$_2$ also shows phase separation in the temperature range 7 - 295 K.

## EXPERIMENTAL DETAILS

The single crystals of Ca(Fe$_{1-x}$Rh$_x$)$_2$As$_2$ ($x = 0$, 0.035, and 0.19) were grown using a self-flux method. Details of the growth of single crystals and their characteristics are given in Ref. 26. The freshly cleaved samples were installed into a He-closed cycle cryostat (Oxford/Cryomech Optistat) with a temperature-controlled range of $T = 7 – 295$ K. Raman-scattering experiments were performed in quasi-backscattering geometry using a $\lambda = 532$ nm solid-state laser. The laser power was set to 7 mW with a spot diameter of approximately 100 μm to avoid heating effects. The spectra were collected via a triple spectrometer (Dilor-XY-500) by a liquid-nitrogen-cooled CCD (Horiba Jobin Yvon, Spectrum One CCD-3000V).

## RESULTS AND DISCUSSION

**Undoped $x = 0$ samples**. CaFe$_2$As$_2$ at ambient pressure and temperature has the paramagnetic tetragonal ThCr$_2$Si$_2$-type structure (space group $I4/mmm$ ($D_{4h}^{17}$, #139) with lattice parameters $a = 3.887(4)$) Å, $c = 11.758(23)$ Å [32]. The crystal structure of CaFe$_2$As$_2$ under ambient conditions is shown in Fig. 1(a). Alkaline-earth ions, Fe and As occupy 2a, 4d, and 4e positions, respectively. According to the factor group analysis for the $I4/mmm$ space group, the total irreducible representation of Raman-active phonon modes is given by four Raman-active phonons: $A_{1g}$(As), $B_{1g}$(Fe), $E_g$(As), and $E_g$(Fe). Upon cooling at ambient pressure, CaFe$_2$As$_2$ undergoes a first order transition at about 170 K from the high temperature tetragonal-paramagnetic phase to the low temperature orthorhombic structure (space group $Fmmm$ ($D_{2h}^{23}$, #69)). It is accompanied by the appearance of a spin density wave (SDW) state, which has been observed in a variety of experiments [33]. This tetragonal-to-orthorhombic (T-O) structural phase transition in a parent, as well as substituted, CaFe$_2$As$_2$ is extremely sensitive to external stresses and strains and can be even partially or completely suppressed. The factor group analysis of low-temperature $Fmmm$ phase with Ca, Fe, and As in 4a, 8f, and 8i Wyckoff positions, respectively yields six Raman-active phonon modes: $\Gamma_{vib} = A_g$(As) + $B_{1g}$(Fe) + $2B_{2g}$(As,Fe) + $2B_{3g}$(As,Fe).

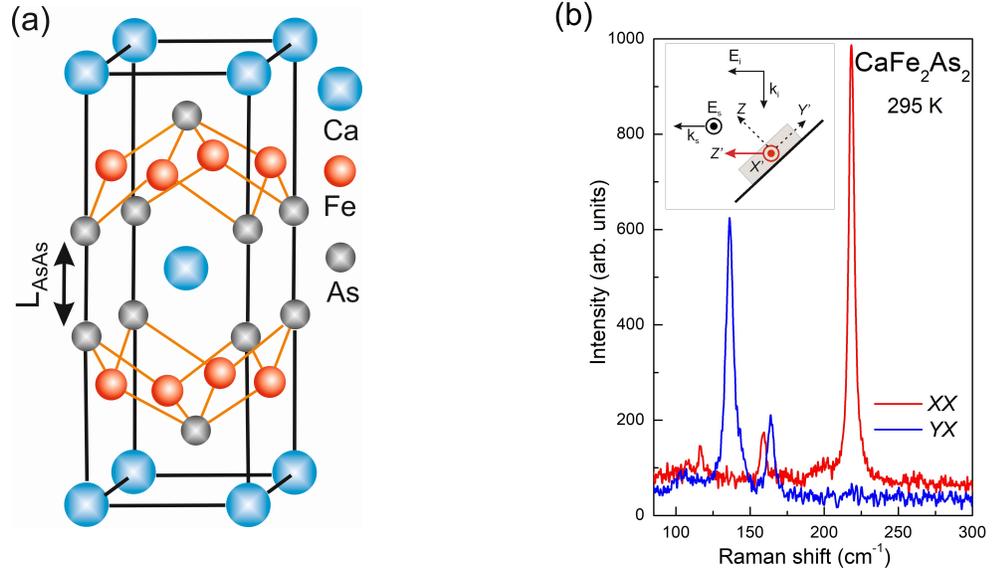

Fig. 1. a) The crystal structure of CaFe$_2$As$_2$ at ambient conditions. b) Room temperature Raman spectra of the CaFe$_2$As$_2$ single crystal measured from the *ab* plane in two scattering geometries. The inset shows the scattering geometry.

K.-Y. Choi and co-workers have reported Raman scattering results on CaFe$_2$As$_2$ single crystals [34]. Two weak phonon modes with energies of 182/189 cm$^{-1}$ ($A_{1g}$) and 204/211 cm$^{-1}$ ($B_{1g}$) were observed in the in-plane scattering geometry at 295/4 K temperatures, respectively. The authors have observed abrupt changes in the temperature-dependent behaviour of the frequencies and linewidths of the $A_{1g}$ and $B_{1g}$ modes upon cooling down through the first-order *I4/mmm* → *Fmmm* structural phase transition near 170 K. Three phonon modes were observed in the electron-doped single crystal Ca(Fe$_{0.95}$Co$_{0.05}$)$_2$As$_2$ [35]. They were assigned to the 205 cm$^{-1}$ - ($B_{1g}$: Fe), 215 cm$^{-1}$ - ($A_{1g}$: As), and 267 cm$^{-1}$ - ($E_g$: Fe and As) phonon modes. All phonon modes showed in the temperature behavior of the parameters (peak position, linewidth measured at half maximum, and integrated intensity) during the passage through the tetragonal-to-orthorhombic structural transition at $T_O$ ~ 140 K.

Simultaneously with these experimental studies, we performed shell-model calculations of the CaFe$_2$As$_2$ lattice dynamics using the general utility lattice program (GULP) [36]. As starting points, model parameters for Fe$^{2+}$, $Ae^{2+}$, and As$^{3-}$ ions were taken from Ref. 37. These model parameters were tuned to achieve reasonable agreement with even-parity phonon modes experimentally observed and calculated earlier in $Ae$Fe$_2$As$_2$. For the homogeneous single-phase CaFe$_2$As$_2$ system, we obtained values of 182 and 201 cm$^{-1}$ for the frequencies of the $A_{1g}$ and $B_{1g}$ modes, respectively. That is in good agreement with early published Raman data [34].

Room temperature Raman spectra of single crystal CaFe$_2$As$_2$ measured from the *ab* plane are shown in Fig. 1(b). The sharpness of the observed peaks is related to the high quality of the studied single crystals. Surprisingly in the room temperature spectra of $CaFe_2As_2$ samples, we observed four well-resolved phonons instead of the two expected. The spectra were fitted using a set of Lorentz functions. The extracted frequencies of the Raman-active optical phonon modes are the two pairs of $A_{1g}$ (159 and 218 cm$^{-1}$) and $B_{1g}$ (136 and 164 cm$^{-1}$) modes. A small leakage of $E_g$(Fe,As) at around 106 cm$^{-1}$ is observed in the spectra due to the quasi-backscattering geometry of the experiment shown in the sketch in the inset of Fig. 1(b). The assignment of the observed Raman active phonon modes to the possible specific symmetry has been performed on the basis of the polarization-dependent Raman tensors with the implementation of the Bilbao Crystallographic Server [38]. Another surprise is that the frequency of the most intensive 218 cm$^{-1}$ $A_{1g}$(As) mode is higher compared to the frequencies of both $B_{1g}$(Fe) modes (which should be the other way around, given the As and Fe ions mass ratio). Moreover, the phonon frequencies of the present investigation do not coincide with those of earlier studies [34,35,37].

The observation of two pairs of Raman bands is evidence for the existence of at least two tetragonal phases in the sample. There is a difficulty with the assignment of the phonon lines observed in the spectra with different phases. If to assume that the ratio of the line intensities corresponds to the ratio of the volumes of the phases present in the sample, then it is reasonable to assign the 218 cm$^{-1}$($A_{1g}$) and 136 cm$^{-1}$($B_{1g}$) to the **t1** tetragonal phase, while the second pair of modes of 159 cm$^{-1}$($A_{1g}$) and 164 cm$^{-1}$($B_{1g}$) - to the **t2** tetragonal phase. With this identification, the frequency order of succession of $A_{1g}$ and $B_{1g}$ phonon modes in phase **t2** coincides with the one observed and calculated earlier [34,37].

The complex phase equilibrium in $CaFe_2As_2$ has been observed and analyzed in many studies [39,40,41,42,43,44]. The general conclusion is that the undoped parent single-crystalline $CaFe_2As_2$ can be stabilized in two slightly different tetragonal phases (PI and PII) after rapid quenching from 850 °C and prolonged annealing at 350 °C, respectively. The PI phase with lattice parameter $c$ = 11.547(1) Å corresponds to a tetragonal structure at room temperature and transforms to a cT phase with $c$ = 10.720(1) Å below the T-cT transition temperature $T_{cT}$ ~100 K. The PII phase also exhibits a tetragonal structure at room temperature but with a slightly longer $c$ ($c$ = 11.763(1) Å). It undergoes a tetragonal-to-orthorhombic (T-O) structural transition at $T_O$ ~170 K. It was found that the PI and PII phases could be reversibly transformed through PI + PII-mixing phase by heat treatment [43]. Various research techniques were used to identify the features of the homogeneity and phase separation in the $CaFe_2As_2$ samples subjected to various heat treatment procedures [39,44,45,46]. Systematic XRD and HRTEM study allowed to determine the micro- and nano-scale structures and their evolution in annealed $CaFe_2As_2$ samples [45]. TEM study of $CaFe_2As_2$ revealed the presence of a clear pseudo-periodic structural modulation at room temperature with a wavelength of around 40 nm and it was interpreted as local structural distortions within the Ca layers [45].

Annealing procedure results in the increasing of the temperature of the antiferromagnetic phase transition. For the most homogeneous crystals, the antiferromagnetic and orthorhombic phase transitions occur at $T_N/T_s$ = 168 K [39]. Our undoped crystals have nearly the same transition temperature which once again emphasizes their high quality.

Raman scattering study and first-principles density functional theory-based calculations were carried out in Ref. 35 to understand the role of magnetic ordering on the phonon subsystem in Co-doped $CaFe_2As_2$. It was found in the theoretical study that the frequency change by a large amount of the order of 20 – 50 cm$^{-1}$ from the nonmagnetic to antiferromagnetic phases indicating the presence of a strong spin–phonon coupling in pure as well as doped $CaFe_2As_2$ systems. The reverse frequency order of the phonon modes $A_{1g}$(As) and $B_{1g}$(Fe) was observed also in the experiment and in calculations which is consistent with our observations for the phase **t1**.

Here it is necessary to dwell on the results of the first-principles calculations, which unravel surprisingly strong interactions between As ions in iron pnictides, the strength of which is controlled by the Fe-spin state [47]. Decreasing the Fe-spin state reduces the Fe-As bonding, which leads to an increase in the As-As bonding and causes the observed huge reduction in lattice parameters. It is noted that this effect is maximal in the case of the $CaFe_2As_2$ system due to the close position of two arsenic ions in adjacent Fe-planes. The second finding in Ref. 47 is the dependence of the $c$-lattice parameter on the iron spin configuration. The authors concluded that this effect is secondary and the main effect is the on-site Fe-spin state.

The impact of the "spin state–lattice" interaction on the phonon subsystem was also studied theoretically in Ref. 48. Taking FeTe compounds as an example, it was found that their phonon spectra undergo significant changes upon the spin state reduction. To describe the properties of iron-containing high-$T_c$ superconductors, the spin state of the iron was suggested to be the control parameter.

The evolution of Raman spectra of undoped $CaFe_2As_2$ with temperature is shown in Fig. 2 in the *XX* and *YX* configurations. Upon cooling, the main effects of the *I*4/*mmm* → *Fmmm* structural transition should be a splitting of the $E_g$ modes and Raman tensor transformation. Indeed, we observe all this during the simultaneous structural T-O phase transition at $T_O$ = 160 K of both tetragonal phases in $CaFe_2As_2$ (see Fig.2). The low-frequency modes of 100-130 cm$^{-1}$ can be assigned with either $B_{2g}$ or $B_{3g}$. We will not dwell on the discussion of these modes.

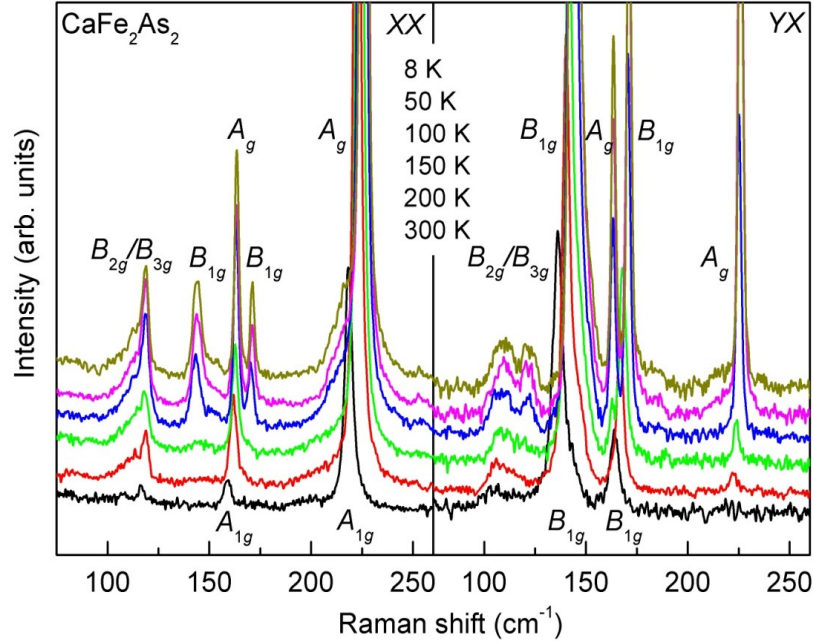

Fig. 2. Temperature dependent Raman spectra of the $CaFe_2As_2$ single crystals measured from the *ab* plane in two scattering geometries. The spectra are vertically shifted for clarity.

Figures 3 (a,c) show temperature evolutions of the peak positions of two groups of Raman modes. Upon lowering temperature all phonon modes show behavior similar to the standard anharmonic model [49] with rather moderate frequency variations and weak anomaly at about $T_O$ = 160 K, related to T-O structural phase transition observed earlier in Raman spectra of $CaFe_2As_2$ [34].

In Figs. 3(b,d) the integrated phonon intensities of the Raman active modes are given. Noteworthy is the change of the slope and significant enhancement of the intensity of the $A_{1g}$(As) and $B_{1g}$(Fe) modes belonging to the **t1** phase at crossing $T_O$ upon cooling. The Raman intensity of a phonon mode consists of the scattering volume multiplied by the electronic polarizability with respect to concerning the particular mode. As the scattering volume and optical penetration depth develop in parallel for **t1** and **t2** phases, anomalies must at least be partly related to the local enhancement of electronic polarizability in the **t1** phase, which is caused by an electronic and structural instability [50] was found from resistivity measurements in $CaFe_2As_2$ [39].

Taking into account the foregoing, we can conclude that already in the parent $CaFe_2As_2$ there is a phase separation into two tetragonal phases. With a high probability, these are the tetragonal collapsed and uncollapsed phases, in which the $Fe^{2+}$ ions are in different spin states. However, both phases undergo the tetra-to-ortho structural phase transition under cooling. This is seen as the appearance of a full set of Raman-allowed phonon lines simultaneously in the *XX* and *YX*-scattering geometries. With our accuracy, we can say about some coincidence of the temperature of the tetra-to-ortho transition for both phases. This can be interpreted in the way that both phases possess magnetic moments. However in an alternative explanation, one can expect that the nonmagnetic cT phase undergoes some orthorhombic distortions from the surrounding magnetic ucT phase under its magnetic ordering.

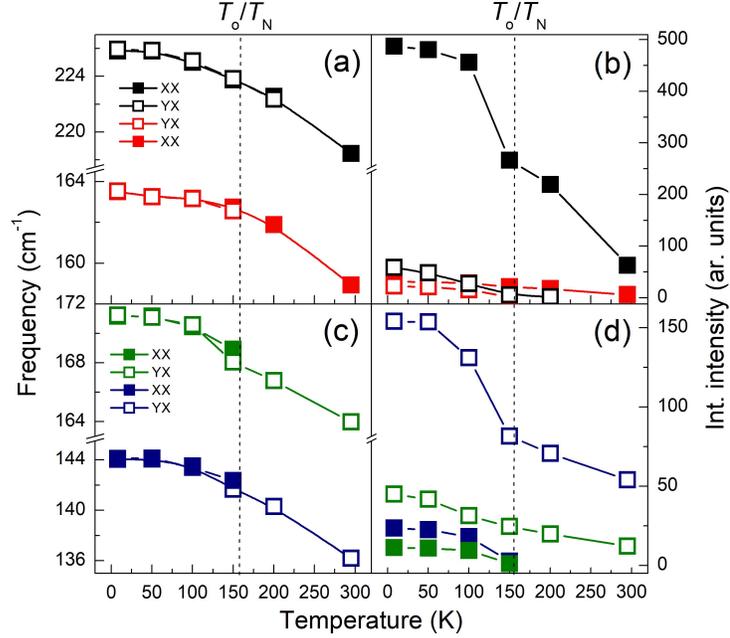

Fig.3. Temperature dependencies of the frequency peak positions (a,c) and integrated intensities (b,d) of the Raman active phonon modes in undoped $CaFe_2As_2$.

**Doped $x = 0.035$ sample.** Now we turn to the analysis of Raman spectra of the doped $Ca(Fe_{0.965}Rh_{0.035})_2As_2$ sample located on the right side close to the verge and in between the orthorhombic AFM phase and tetragonal nonmagnetic phase and to the superconducting phase of $Ca(Fe_{1-x}Rh_x)_2As_2$ [26]. Figure 4 shows the temperature-dependent evolution of polarized Raman spectra of the $Ca(Fe_{0.965}Rh_{0.035})_2As_2$ single crystal. Room-temperature spectra of the $x = 0.035$ Rh-doped sample are similar in the number of strong phonon lines and their frequency positions to these of the undoped sample (Fig.2). So, the same two-phase separation scenario exists for the $x = 0.035$ Rh-doped sample. It should be noted that small orthorhombic distortions are already present in the tetragonal phase $\tau 1$ at room temperature, which is seen due to the presence in the spectra of low-intensity $A_{1g}(As)$ and $B_{1g}(Fe)$ lines in the forbidden scattering geometries. The reason for this may be the internal pressure induced by Rh doping. However, at cooling down to $T^* \approx 100$ K not any changes are observed in the spectra similar to the changes in undoped samples during the tetra-to-ortho structural phase transition. Here we can state some preliminary conclusions. The nonmagnetic uncollapsed phase does not undergo the tetra-ortho structural phase transition up to 100 K. Therefore, such a transition, if it exists, should have a magnetic origin.

Below the critical temperature $T^*$, we observe the appearance of a number of new phonon lines in the spectra (see Figs. 4,5). This $T$-dependence completely excluded the Rh-impurity origin of the new lines. To be sure that our "unusual" observations are not connected with the resonance effect, we carried Raman measurements with different excitation energies. The results, which unequivocally rule out the presence of resonance effects, are shown in Fig. 6. It is clearly seen that Raman spectra measured at different laser excitation frequencies practically coincide.

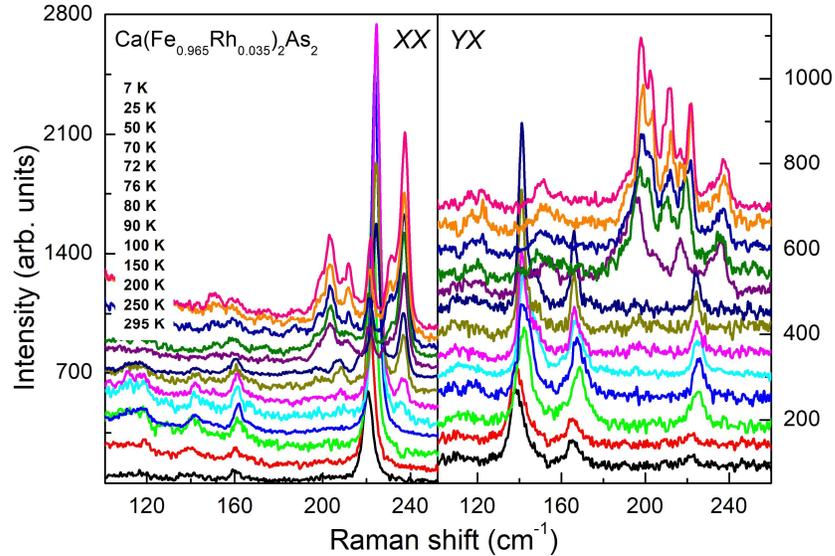

Fig. 4. Temperature dependent $XX$ and $YX$ Raman spectra of the Ca(Fe$_{0.965}$Rh$_{0.035}$)$_2$As$_2$ single crystal.

Figure 7 shows the temperature behavior of selected phonon line parameters (frequency, linewidth, and integrated intensity) in the frequency region of 220 – 240 cm$^{-1}$ together with the temperature dependence of magnetic susceptibility, $M/H$, taken from Ref. 26. Observations from both studies clearly indicate that there is an intermediate region about 28 K wide between $T^*$ and $T_{cT}$. Note that the coexistence of several CaFe$_2$As$_2$ known phases was observed in a temperature window between 45 K and 90 K in the samples with the highly inhomogeneous strain fields which led both to the formation of domains of varying $c$-lattice parameters and to static random displacements of the atoms from their equilibrium positions [39].

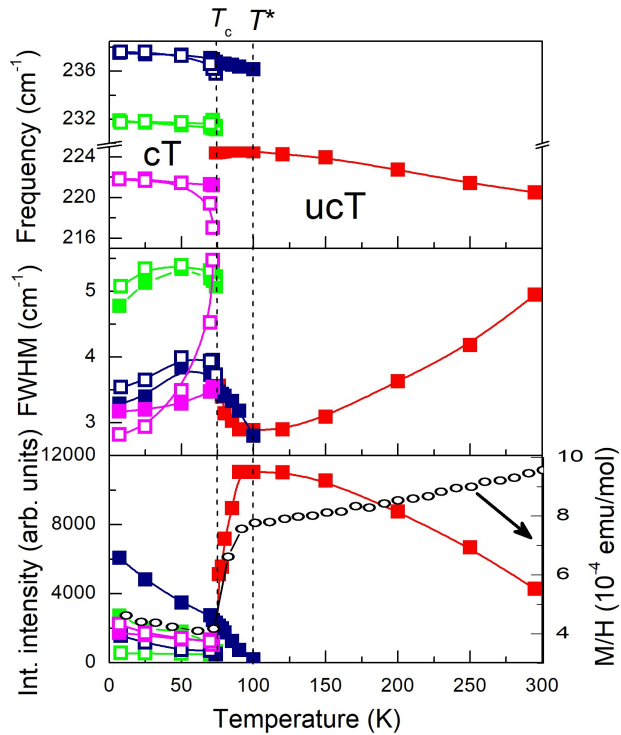

Fig. 5. Temperature behavior of the phonon lines parameters (frequency, linewidth, and integrated intensity) in the frequency region of 215 – 240 cm$^{-1}$ together with the temperature dependence of magnetic susceptibility, *M*/*H*, taken from Ref. 26.

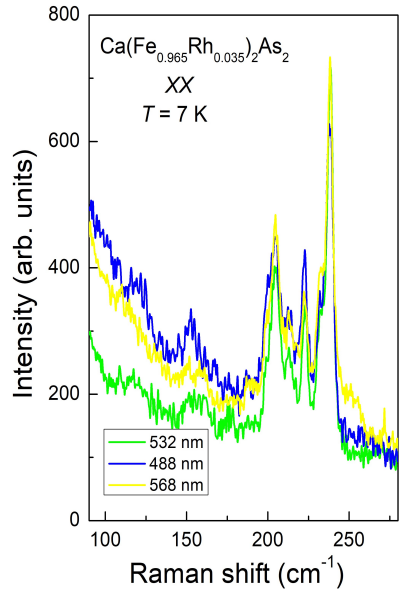

Fig. 6. Raman spectra of Ca(Fe$_{0.965}$Rh$_{0.035}$)$_2$As$_2$ measured using three different wavelength and excitation energies of the incident Laser radiation.

A quite unexpected result is stated with cooling below $T_{cT}$ = 72 K. From general considerations, earlier X-ray and neutron scattering experiments, and according to the results of nonmagnetic DFT calculations [25], it follows that the vibrational properties of CaFe$_2$As$_2$ change surprisingly little through uncollapsed → collapsed tetragonal phase transition, except for a few modes, which are very sensitive to the *c*-axis lattice parameter. Indeed, the changes in the spectra of the **t2** phase are in very good agreement with the results of studies in Ref. 25, namely, a noticeable frequency shift of the $A_{1g}$(161 cm$^{-1}$) and $B_{1g}$(167 cm$^{-1}$) lines is observed below the temperature $T_{cT}$ of ucT→ cT phase transition. As for the **t1** phase, the phonon Raman spectra at low temperatures change completely - phonon lines of the uncollapsed tetragonal phase fully disappear in the spectra and new lines appear. Moreover, below $T_{cT}$ all phonon lines are active in both parallel and crossed scattering geometries. The only explanation for this feature is the emergence of new order parameters with new translation symmetry.

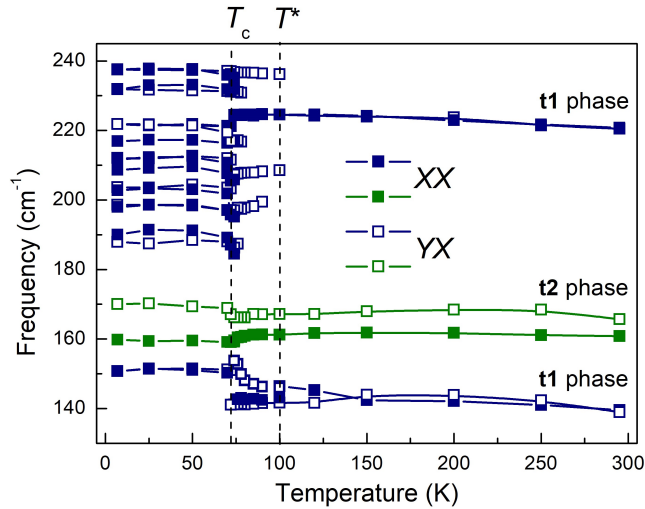

Fig. 7. Temperature dependence of frequencies of all phonon lines observed in the frequency region of 130 – 245 cm$^{-1}$ in Raman spectra of Ca(Fe$_{0.965}$Rh$_{0.035}$)$_2$As$_2$ measured in the *XX* and *YX* scattering geometries.

The Fe-magnetism is required for an unconventional mechanism of high-$T_c$ superconductivity in the iron pnictides. The competition of superconductivity and magnetism is especially pronounced in the re-entrance of the SDW phases $C_2$ twofold and $C_4$ fourfold rotational symmetry inside the superconducting dome both in electron- [28] and hole-doped [29,30] *Ae*Fe$_2$As$_2$ compounds. Numerous studies, including published results in Ref. 47, showed also that the Fe-magnetism is not totally lost in the cT-phase, but it is partially reduced. Different mechanisms, namely, magnetic and orbital ordering, for symmetry lowering in the orthorhombic and the re-entrant tetragonal phases in the hole-doped Ba$_{0.76}$Na$_{0.24}$Fe$_2$As$_2$ were considered in detail in Ref. [51]. Note that in this compound the transition is not complete and the re-entrant phase coexists with the orthorhombic one down to the lowest measured temperature 1.5 K. The magnetic mechanism for symmetry lowering in the re-entrant tetragonal phase with out-of-plane magnetic moments implies a two-***k*** magnetic structure with tetragonal $P_C4_2/ncm$ symmetry imposes zero dipole magnetic moments for half of the Fe sites. In this case and the crystal structure symmetry (without magnetic subsystem) of the system is well approximated by the parent $I4/mmm$ space group. The orbital ordering mechanism predicts the crystal structure symmetry lowering down to $P4/mnc$ or $I422$ depending on the stacking of the ($ab$) ordered layers along the $c$ axis. Both types of orbital ordering do not allow any atomic displacements in comparison with the parent $I4/mmm$ space group. Combination of the magnetic order with orbital $M_3^+/\Gamma_1^-$ ordering results in the orthorhombic $P_Cccn/C_A222_1$ magnetic space groups, respectively.

In Ref. 51 the most preferable magnetic structure in the re-entrant phase of the Ba$_{0.76}$Na$_{0.24}$Fe$_2$As$_2$ is the tetragonal one, $P_C4_2/ncm$ with G-type order of nonmagnetic and magnetic states on neighboring iron atoms (see Fig. 4 in Ref. 51). Such a structure should lead to the appearance of a number of new phonon modes in our Raman spectra. Nevertheless, with a high probability, we can assert a novel re-entrance of the $C_2$-symmetric phase in $x = 0.035$ doped sample at $T < 72$ K that coexists with the collapsed tetragonal phase. The last statement is based on the fact that: i) insignificant orthorhombic distortions present in the Raman spectra in the entire temperature range starting from the room temperature; ii) at temperatures below $T_c = 72$ K, all appeared phonon excitations are active in both parallel and perpendicular scattering geometries. Note that these new phonon modes do not present in the spectra of the magnetically ordered uniform phase of the $x = 0$ undoped sample. It evidences in favor of a magnetic structure distinctive from the usual orthorhombic one which appears at low temperatures in the $x = 0.035$ sample. A precise structural determination in a doped 122-$Ae$Fe$_2$As$_2$ system in a wide temperature and concentration range is a challenge that can be solved with modern diffraction facilities [51].

**Overdoped $x = 0.19$ sample.** Ca(Fe$_{0.81}$Rh$_{0.19}$)$_2$As$_2$ undergoes a first-order phase transition from the ucT phase to the cT at the temperature $T_{cT} = 295$ K [26]. Raman spectra of Ca(Fe$_{0.81}$Rh$_{0.19}$)$_2$As$_2$ at 295 and 420 K are shown in Fig. 8. At these temperatures, the observed Raman spectra are very weak, the phonon lines are broadened and pronounced quasielastic electronic background is observed due to the Rh-doping. The latter result in one extra electron per substituted atom. The inset in Fig. 8 shows a Raman spectrum of Ca(Fe$_{0.81}$Rh$_{0.19}$)$_2$As$_2$ at $T = 295$ K with a subtracted quasielastic background. Impurities induce the appearance of additional bands in the frequency region of high (245 cm$^{-1}$) and low (165 cm$^{-1}$) energy phonons clearly seen in the spectrum.

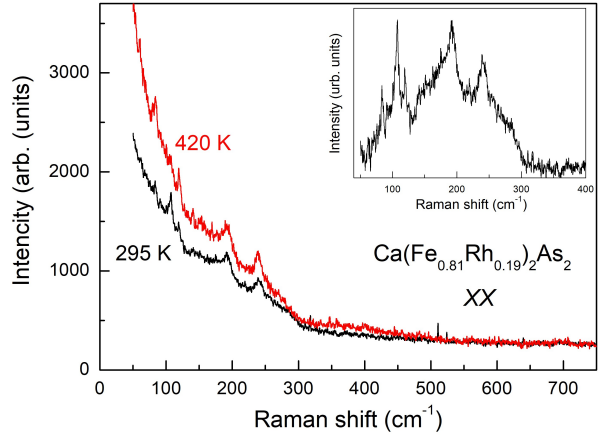

Fig. 8. Raman spectra of the overdoped single crystal of Ca(Fe$_{0.81}$Rh$_{0.19}$)$_2$As$_2$ measured from the *ab* plane in *XX* scattering geometry at 295 and 420 K. The inset shows a spectrum at $T = 295$ K with subtracted quasielastic background.

Room temperature Raman measurements performed using a micro-Raman setup with a much higher power density of the excitation laser yielded spectra containing broad phonon lines in the same frequency region where intensive phonon lines are observed in the cT phase of the $x = 0.035$ sample (Fig. 9). This indicates that the $x = 0.19$ sample, located on the right side of the electronic phase diagram of Ca(Fe$_{1-x}$Rh$_x$)$_2$As$_2$, shows phase separation at high temperatures, similar to $x = 0.035$ sample at $T_c < 72$ K.

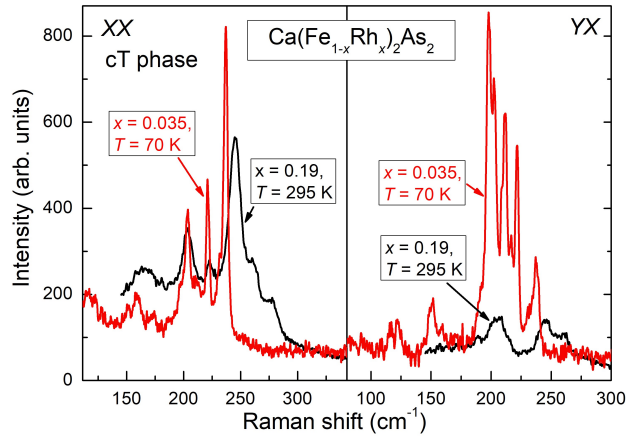

Fig. 9. Comparison of room temperature Raman spectra of Ca(Fe$_{1-x}$Rh$_x$)$_2$As$_2$, $x = 0.19$, with 70 K spectra of $x = 0.035$ in the cT phase. Data from the *ab* plane and in *XX* and *YX* polarization, respectively.

## CONCLUSIONS

Since Fe-based superconductors are distinct from the other families, their unique physics should be connected with specific features of the Fe ion itself. In fact, Fe$^{2+}$ is famous for its spin crossover involving $S = 0, 1, 2$ states with different ionic radii. In an extended solid they can be used as a concept to describe properties of a many body ground states that are otherwise difficult to relate. The $S = 0, 1$ states are most important for the model of magnetism in iron pnictides since they can overlap in the many-body ground state by an inter-ion exchange of just two electrons. The spin-state flexibility of $Fe^{2+}$ can be tuned by external or chemical pressure.

Since unconventional superconductivity implies Fe-magnetism, detailed knowledge of the phase diagram and the nature of the competing magnetic and superconducting phases are imperative for a deeper understanding of the physics of iron-based superconductivity. Motivated by this we apply Raman spectroscopy to study three samples with $x = 0$, 0.035, and 0.19 which represent important parts of the $Ca(Fe_{1-x}Rh_x)_2As_2$ phase diagram. Several unexpected and very interesting results have been obtained. Among them:

i) Different spectra identified with two tetragonal phases were observed for the first time in undoped (x = 0) and the moderately doped ($x = 0.035$) samples at room temperature. We attribute these phases with different volumes of the unit cells to the collapsed and the uncollapsed tetragonal phases contain iron ions with different spin states, $S = 0$ and $S = 1$. This coexistence is present on a larger crystalline length scale and is not limited to local distortions.

ii) Within our accuracy, we can state that both phases in undoped samples undergo tetra-to-ortho phase transitions approximately at the same temperature of the structural and magnetic transitions, $T_c/T_N$. This is interpreted in the sense that both cT and ucT phases simultaneously possess magnetic moments. However, as an alternative we also consider that the nonmagnetic cT phase undergoes some orthorhombic distortions from the surrounding magnetic ucT phase under its magnetic ordering.

iii) With cooling down to $T^* \approx 100$ K no changes are observed in the spectra of the doped sample that could be compared with effects in the undoped sample during the tetra-to-ortho structural phase transition. This means that the uncollapsed tetragonal phase does not undergo a tetra-to-ortho structural phase transition up to 100 K. Therefore, any related transition, if it exists, should have a magnetic origin.

iv) For temperatures below $T^*$ we observe the emergence of new phonon lines in the spectra. The only explanation for this feature is the condensation of an order parameters with a new translation symmetry in the system. At further cooling through the ucT–cT transition temperature, $T_{cT} = 72$ K [26], drastic changes occur in the spectra. According to our analysis, uncollapsed → collapsed tetragonal phase transition is successfully carried out for the former ucT phase. Simultaneously, many new phonon lines in both *XX* and *YX* scattering geometries appear in the spectra. With a high probability, we can assert a novel re-entrance of the $C_2$-symmetric phase distinctive from the usual orthorhombic twofold one in undoped samples.

v) The overdoped sample with $x = 0.19$ located on the right side of the phase diagram [26] also shows phase separation with Raman spectra similar to the ones of the doped sample, measured at $T < T_c$.

Further studies are required to accurately establish the symmetry of the newly discovered phases in Ca(Fe$_{1-x}$Rh$_x$)$_2$As$_2$ and we refer to Ref. 51 for possible scenarios of magnetic and orbital ordering. Possible space groups could be distinguished in resonant X-ray Bragg diffraction patterns created by Templeton & Templeton scattering.

## ACKNOWLEDGMENTS

A. Yu. Glamazda thanks Dr. A. V. Peschanskii for fruitful discussions. We thank the VW Foundation for support within A139645. Furthermore, we acknowledge support by EXC-2123 Quantum Frontiers – 390837967.